# Designing for Healthy, Affordable, and Sustainable Human-HVAC Interactions for Heating in Smart Homes

Delong Korus-Du*
delong.korusdu@uni-siegen.de
Verbraucherinformatik, Department of Human-Computer Interactions, University of Siegen
Siegen, Germany

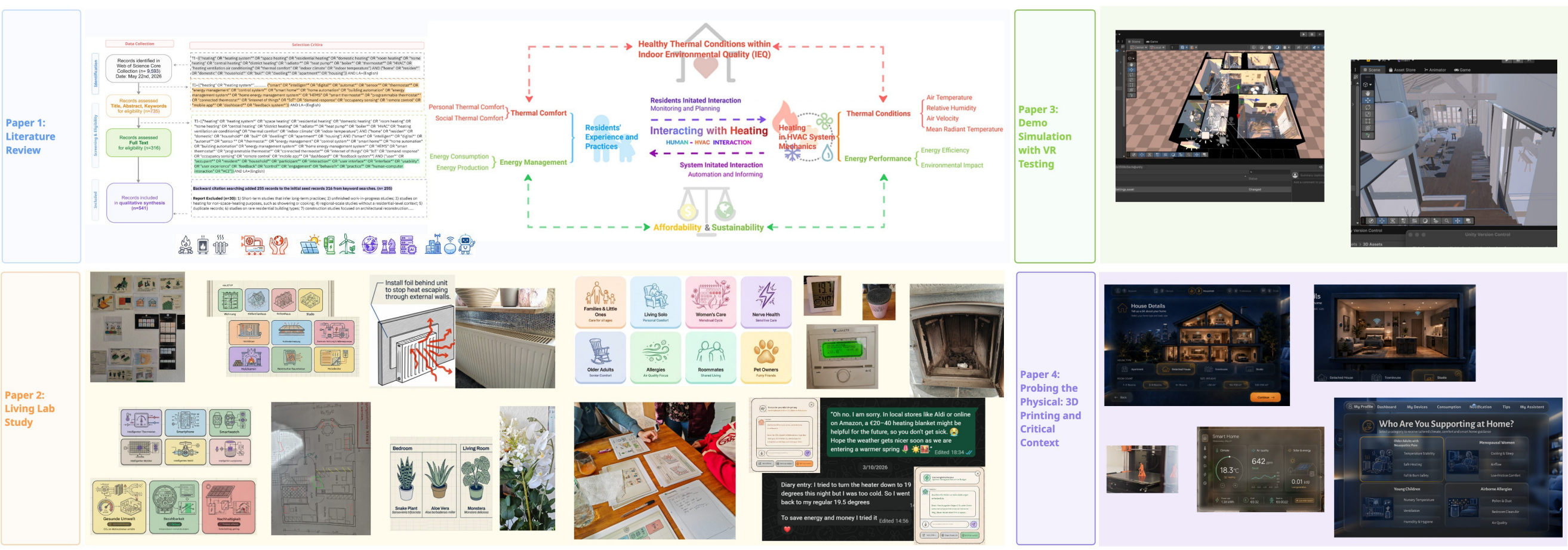

**Figure 1: Overview of the doctoral research programme. Preliminary results include a multidisciplinary literature review and a longitudinal living lab study of domestic heating practices. Expected next steps include a virtual-reality simulation study and the development and critical evaluation of a physical 3D-printed smart-home interface. The four studies collectively investigate thermal comfort, indoor environmental quality, heating practices, energy performance, affordability, and sustainability.**

## Abstract

As geopolitical tensions, energy crises, and energy-intensive AI infrastructure intensify concerns about demand, affordability, and resilience, communities increasingly encounter these challenges through everyday energy practices, particularly winter heating. Against this background, the doctoral exposé, *"Designing Human-HVAC Interaction for Healthy, Affordable, and Sustainable Heating in Smart Homes"*, is structured around four chapters.

First, a multidisciplinary literature review defines and positions Human-HVAC Interaction, focusing on heating in smart homes. Second, longitudinal living lab studies with design probes examine everyday heating practices, thermal comfort, and indoor environmental quality, with attention to thermally vulnerable groups such as older adults, pregnant or menopausal women, parents with infants, and people affected by allergies or airborne pollutants. Third, a VR-based smart home demonstrator explores how heating and IEQ scenarios can be prototyped and evaluated as a virtual living lab, while critically examining the limits of representing bodily indoor climate conditions through VR. Fourth, follow-up design studies examine how VR-based insights can be translated into physical-digital prototypes that combine digital fabrication, distributed environmental sensing, and diverse interface forms for critical heating and IEQ contexts.

The thesis aims to contribute a design-oriented understanding of Human-HVAC Interaction by building from a multidisciplinary literature review to empirical living lab and co-design studies, VR-based prototyping, and physical system development, examining how smart home users make sense of, negotiate, and respond to smart HVAC system.

# 1 Introduction

Residential heating is a foundational infrastructure for habitation in cold and temperate climates, yet it is increasingly becoming a site of social, environmental, and economic tension. Buildings in polar, subpolar, and temperate zones account for about 30% of total final energy use, with nearly half devoted to heating consumption [15, 16, 19, 62]. Since the mid-twentieth century, people have also spent more time indoors [29], making residential HVAC systems central to everyday comfort, indoor environmental quality, household expenditure, and environmental impact [35, 46]. These pressures are expected to intensify. In Germany and across Europe, recent energy price shocks have burdened private households and affected industrial competitiveness; at the same time, expanding AI and data-centre infrastructure is expected to add further electricity demand, as reported by the *Bundesministerium für Wirtschaft und Energie* (Federal Ministry for Economic Affairs and Energy, BMWK) [18, 54].

At household level, the energy crisis has turned heating from a routine background service into a recurring negotiation over which rooms to heat, when to ventilate, and how to balance bills against health and care needs [6, 58]. These pressures are uneven: low-income households, residents of inefficient buildings, tenants with limited control, and thermally vulnerable occupants have less capacity to respond, while already constrained consumption can make further reductions unsafe or infeasible [5, 37, 56].

Existing research has made significant progress in improving the efficiency and performance of heating systems. Thermal comfort and HVAC engineering research has developed predictive models and control algorithms based on thermophysical and physiological parameters [17, 23, 36, 62]. Social practice research has shown that heating consumption is shaped by everyday routines, social norms, material infrastructures, competences, and meanings [20, 34, 48, 50]. Sustainable HCI has further argued that domestic energy technologies should not be treated as neutral optimization tools, but as socio-technical systems embedded in everyday life [4, 55, 64]. However, research on how residents interact with HVAC systems, particularly heating, remains fragmented across HCI, building science, engineering, architecture, environmental psychology, sociology, and energy studies [11, 28, 44].

This fragmentation becomes visible in the design of smart home energy management systems. Contemporary systems often emphasize monitoring, prediction, automation, and feedback [2, 38, 40], but the everyday use of heating is shaped by conditions that cannot be addressed through information or automation alone [9, 10]. Households interpret and act on heating information through uneven infrastructures [49], limited budgets, health vulnerabilities [3, 27, 43], caregiving responsibilities [39], and different levels of practical knowledge [26, 47, 51, 53, 57]. Asking residents simply to reduce energy use risks overlooking the conditions that make reduction possible, safe, fair, or desirable [52, 56]. Over-reliance on automation can also increase domestic care work, disrupt routines, trigger information avoidance, or lead to the abandonment of monitoring practices over time [24, 47]. These challenges are especially important for thermally and environmentally vulnerable groups. Older adults, people with cold-related pain, women during pregnancy or menopause, parents with infants or small children, and people affected by allergies or airborne pollutants may experience heating, ventilation, humidity, and air quality as matters of health, care, and safety rather than only comfort or efficiency [27, 41–43, 60, 63]. Sustainable Human-HVAC Interaction therefore requires more than efficient automation or persuasive feedback. It requires interfaces and learning mechanisms that help residents interpret IEQ, understand system behaviour, develop practical competencies, and make situated decisions under infrastructural, financial, bodily, and environmental constraints [8, 14, 59].

This thesis responds to these challenges through a stepwise research trajectory. The first chapter develops an interdisciplinary literature review to clarify and position Human-HVAC Interaction. The second chapter uses longitudinal living lab and co-design studies to examine everyday heating practices, thermal comfort, and IEQ in domestic settings, with attention to vulnerable groups [1, 21, 25]. Building on these completed studies, the second chapter is to investigate how design knowledge from real-life settings can be translated into prototyping environments. However, as longitudinal living lab studies are valuable, they require substantial time, resources, and technical infrastructure; on the other hand, low-fidelity prototypes or existing commercial products may not fully demonstrate future smart heating interactions, while real-world deployment makes it difficult to rapidly vary household settings, system behaviours, or critical IEQ scenarios. Thus, in the third chapter, virtual reality demos will be developed to function as a virtual living lab for prototyping and studying Human-HVAC Interaction before physical implementation. At the same time, VR must be treated critically, because indoor climate is not only visual or informational but also bodily and sensory. Temperature, air quality, humidity, and airflow cannot be fully reproduced through visual simulation alone [13, 22, 30, 61]. Therefore, in the fourth chapter, various physical artifacts will be made and deployed in the real world, based on the research groundings built upon the previous chapters.

Building on this tension between situated domestic practice, virtual prototyping, and physical implementation, this PhD investigates how Human-HVAC Interaction can be conceptualized, designed, prototyped, and translated into physical device-based systems to support healthy IEQ, affordable energy use, and sustainable heating practices. Across four connected chapters, the dissertation moves from conceptual clarification to empirical living lab and co-design studies, VR-based prototyping, and design exploration for critical IEQ contexts in smart homes and other vulnerable indoor environments.

# 2 Research Questions

The overarching research question is:

**How can Human-HVAC Interaction be designed to support healthy indoor environmental quality, affordable domestic energy use, and sustainable heating practices in smart homes?**

This question is addressed through four sub-questions:

(1) **Conceptualizing Human-HVAC Interaction:** How are Human-HVAC Interaction and heating-related practices conceptualized and studied across HCI, building science, energy research, and related disciplines?

(2) **Understanding and supporting domestic heating competencies:** How can smart home energy management interfaces support households in developing situated, feasible, and health-sensitive competencies for heating, IEQ, and energy management?
(3) **Prototyping Human-HVAC Interaction through virtual living labs:** How can virtual reality be used to prototype, demonstrate, and study smart heating and IEQ scenarios with the focus on empathic cohabitation with others, and what are its methodological limits for representing bodily and sensory indoor climate conditions?
(4) **Translating Human-HVAC Interaction into physical-digital prototypes:** How can insights from living lab and VR-based studies be translated into digitally fabricated, sensor-based, and networked prototypes across tangible, wearable, ambient, and voice-based interfaces to support residents in interpreting and responding to heating, energy, and IEQ conditions?

## 3 Preliminary Results

As shown in the Teaser Figure on the first page, the PhD has established its conceptual and empirical foundation through the first two studies.

**Chapter I** has been completed and is currently under review. This study is a multidisciplinary literature review of 541 papers that conceptualizes Human-HVAC Interaction, with a particular focus on heating. The review brings together work from HCI, building science, energy research, architecture, engineering, physiology, psychology, sociology, and design. It provides the conceptual foundation for the dissertation by framing Human-HVAC Interaction as situated interaction dynamics between residents' experiences and practices and HVAC system mechanics. Shown in the Teaser Figure on the first page, our literature review study conceptualizes Human–HVAC Interaction (H-HVAC) as the situated dynamics of control and feedback between users and systems. User-initiated interactions involve monitoring past and present HVAC performance and planning future operation, while system-initiated interactions use sensor networks to trigger automation or guide user action. These dynamics connect residents' comfort and energy management with HVAC mechanics' thermal conditions and energy performance [33].

**Chapter II** has collected empirical and design-oriented insights into how vulnerable users and household members can better understand and respond to heating-related needs. The study focuses on how households develop practical competencies for care, comfort, affordability, and IEQ. For example, the findings show that household members may need to develop shared competencies to understand each other's thermal needs, negotiate heating practices, and identify affordable and accessible interventions. These include low-cost embodied and spatial actions such as placing reflective foil behind radiators, adjusting ventilation routines, moving furniture away from cold walls, using localized warming, or making small infrastructural changes based on situated feedback. The current analysis organizes these findings into design implications and a semantic structure of learning mechanisms for smart home energy management interfaces. The results show that thermal vulnerable groups already have high competency in understanding their needs, but interestingly, their cohabitants may not be fully aware. Thus, besides the general recommended tips that target at infrastructural and urban aspects, user specific recommendations are mostly needed for their cohabitants than the thermal vulnerable groups themselves [31].

## 4 Expected Next Steps

**Chapter III** builds on the findings of the living lab and co-design work from the second study by developing a VR-based smart home environment for prototyping and evaluating Human-HVAC Interaction scenarios. The environment translates empirical findings into interactive demonstrations, including smart displays, sensor-based feedback, heating-related recommendations, and IEQ scenarios. The study examines whether VR can function as a virtual living lab for exploring household heating practices, HVAC interaction, and learning mechanisms for sustainable energy management. Moreover, as VR being argued as the "ultimate empathy machine", this approach can allow users play in the role that they are not familiar with - such as a husband playing as his wife experiencing menstrual hot flash, or caregivers as older adults. However, this VR simulation approach also needs be critically examined regarding the methodological limits of VR, since temperature, air quality, humidity, and airflow are bodily and sensory conditions that cannot be fully reproduced through visual simulation alone [32].

**Chapter IV** will translate selected VR-based scenarios into physical-digital prototypes, supported by digital fabrication facilities such as the Fab Lab at the University of Siegen. This may include digitally fabricated devices, distributed environmental sensor networks, and interaction concepts across tangible interfaces, wearable technologies, smart glasses, voice assistants, and ambient feedback systems [7, 12, 45, 65]. Rather than developing a complete commercial HVAC system, the study examines how residents might interpret and respond to critical heating, energy, and IEQ conditions through physical, networked, and embodied forms of interaction. Beyond domestic smart homes, this may inform future work in more critical indoor contexts, such as hospitals, operating rooms, older care, neonatal care, and clean rooms, where indoor climate conditions are closely tied to health, safety, and care, as well as cooling in Human-HVAC interaction.

## 5 Expected Contributions

This PhD contributes to Sustainable HCI, human-building interaction, and smart home research by conceptualizing Human-HVAC Interaction through a multidisciplinary literature review; examining how households interpret, negotiate, and respond to heating, energy, and IEQ recommendations through living lab and co-design studies; and exploring how these insights can be prototyped, evaluated, and translated through VR-based smart home scenarios and physical-digital prototypes.

## Acknowledgments

This project has received funding from the European Union's Horizon 2020 research and innovation programme under the Marie

Skłodowska-Curie grant agreement No. 955422, and was also supported by the FUSION project, funded by the German Federal Ministry of Education and Research (BMBF) programme *Innovative University*.